
\magnification=\magstep1
\parskip=.2truecm

\font\ti=cmbx10 scaled\magstep1
\font\eightrm=cmr8
\font\ninerm=cmr9
\def\br{\hfill\break\noindent}
\def\deg{{\rm deg}}

\def \ot {\otimes}
\def \g5{\gamma_5}

\def \l4{\Bigl( {\rm Tr}( KK^*)^2-({\rm Tr}KK^*)^2\Bigr)}

\def \k2{{\rm Tr}KK^*}
\def \slash#1{/\kern -6pt#1}

\def\inbar{\,\vrule height1.5ex width.4pt depth0pt}
\font\cmss=cmss10 \font\cmsss=cmss10 at 7pt
\def\C{\relax\hbox{$\inbar\kern-.3em{\rm C}$}}
\def\Z{\relax\ifmmode\mathchoice
{\hbox{\cmss Z\kern-.4em Z}}{\hbox{\cmss Z\kern-.4em Z}}
{\lower.9pt\hbox{\cmsss Z\kern-.4em Z}}
{\lower1.2pt\hbox{\cmsss Z\kern-.4em Z}}\else{\cmss Z\kern-.4em Z}\fi}
\def\Q{\relax\hbox{$\inbar\kern-.3em{\rm Q}$}}
\def\R{\relax{\rm I\kern-.18em R}}

\def\End{{\rm End}}
\def\Ker{{\rm Ker}}

\pageno=0

\def\tr{{\rm tr}}
%
%
\baselineskip=.5truecm
\footline={\hfill}
{\hfill ETH-TH/1992-18}
\vskip.2truecm
\vskip2.1truecm
\centerline{\ti Gravity in Non-Commutative Geometry }
\vskip1.2truecm
\centerline{  A. H. Chamseddine$^{1,}$\footnote*
{\ninerm Supported by the Swiss National Foundation (SNF)},
G. Felder$^{2}$ and J. Fr\"ohlich$^{3}$ }
\vskip.8truecm
\centerline{$^{1}$ Theoretische Physik, Universit\"at Z\"urich, CH-%
8001 Z\"urich, Switzerland}
\centerline{$^{2}$ Mathematik, ETH-Zentrum, CH-8092 Z\"urich, Switzerland}
\centerline{$^{3}$ Theoretische Physik, ETH-H\"onggerberg,
 CH-8093 Z\"urich, Switzerland}
\vskip1.2truecm

\centerline{\bf Abstract}
\vskip.5truecm

We study general relativity in the framework of non-commutative
differential geometry. In particular, we introduce a gravity
action for a space-time which is the product of a four dimensional
manifold by a two-point space. In the simplest situation, where
the Riemannian metric is taken to be the same on the two copies
of the manifold, one obtains a model of a scalar field coupled
to Einstein gravity. This field is geometrically interpreted
as describing the distance between the two points in the
internal space.

\eject
\baselineskip=.6truecm
\footline={\hss\eightrm\folio\hss}

\centerline{}
\vskip1.1truecm
{\bf\noindent 1. Introduction}
\vskip.2truecm

\noindent
The poor understanding we have of physics at
very short distances might lead to expect that our description
of space-time at tiny distances is inadequate. No convincing
alternative description is known, but different routes to progress
have been proposed. One such proposal is to try to
formulate physics on some non-commutative space-time.
There appear to be too many possibilities to do this, and
it is difficult to see what the right choice is. So
the strategy is to consider slight variations of
commutative geometry, and to see whether reasonable models
can be constructed. This is the approach followed by
Connes [1], and Connes and Lott [2,3]. They consider
a model of commutative geometry (a Kaluza-Klein theory
with an internal space consisting of two points), but
use non-commutative geometry to define metric properties.
The result is an economical way of deriving the standard
model in which, roughly speaking, the Higgs field appears as
the component of the gauge field in the internal direction.

In this paper, we show how gravity, in its simplest form,
can be introduced in  this context. We first
propose a generalization of the basic notions
of Riemannian geometry. This construction is
based on the definition of the Riemannian metric
as an inner product on cotangent space.
Our definition differs from the one advocated
by Connes, who proposes to replace the notion
of Riemannian metric by the notion of $K$-cycle.
We show however that, for the class of (commutative)
Kaluza-Klein
models we consider, the two approaches can
be related, and each $K$-cycle gives
rise to a Riemannian metric in our sense.
After this, we propose a generalized Einstein-Hilbert
action and see how it looks like in the
case of a Kaluza-Klein model with a two-point internal
space.

The construction illustrates an interesting feature
of non-commutative geometry for commutative spaces:
the fact that the metric structure is more general
allows one to consider a class of metric spaces more
general than Riemannian manifolds, in which however
differential geometric notions, such as connections
and curvature, still make sense.

The physical picture emerging from this is of
a gravitational field described by a Riemannian
metric on a four-dimensional space time plus
a scalar field which encodes the distance between
the two points in the internal space.
This field is massless and couples in a minimal
way to gravity.
Its vacuum expectation value turns out to determine the
scale of weak interactions in the formalism of [3].

\vskip 1truecm

{\bf \noindent 2. Riemannian geometry}
\vskip .2truecm
\noindent
In this section we develop some concepts of Riemannian geometry
in the more general context of non-commutative spaces.
Let $\Omega^\cdot$ be a $\Z$-graded differential
algebra over $\R$ or $\C$. This means that
$\Omega^\cdot=\oplus_0^\infty\Omega^n$ is a graded
complex of vector spaces with differential $d:\Omega^n\to\Omega^{n+1}$
and that there is an associative product
$m:\Omega^n\otimes\Omega^m\to \Omega^{n+m}$. In particular,
$A=\Omega^0$ is an algebra, and $\Omega^n$ is a two sided
$A$ module. We will always assume that $\Omega^\cdot$ has
a unit $1\in A$.
The algebra $A$ is to be thought of as a generalization of
the algebra of functions on a manifold, and $\Omega^\cdot$ as a
generalization of the space of differential forms.
The most important example for us is
Connes' algebra of universal forms $\Omega^\cdot(A)$ over
an algebra $A$.
 It is generated by symbols $f$, of degree zero, and $df$,
of degree one, $f\in A$,
with relations $d(fg)=df\,g+f\,dg$, $f$, $g\in A$, and $d\,1=0$.
The notation is consistent since $\Omega^0(A)=A$.

To do Riemannian geometry we need a notion of Levi-Civita
connection.

In general, a {\it connection\/} on a left $A$ module $E$ is,
 by definition,
a linear map $\nabla:E\to\Omega^1\ot_A E$ such that for
any $f\in A$ and $s\in E$,
$$
\nabla (fs)=df\ot s+f\nabla s.
$$
For any left $A$ module $E$, define $\Omega^\cdot E$
to be the graded left $\Omega^\cdot$ module, $\Omega^\cdot E=
\Omega^\cdot\ot_AE$, of ``$E$-valued differential forms''. A
connection $\nabla$ on $E$ extends uniquely to a linear
map of degree one $\nabla: \Omega^\cdot E\to\Omega^\cdot E$
with the property that, for any homogeneous $\alpha\in \Omega^\cdot$,
$\phi\in\Omega^\cdot E$,
$$
\nabla(\alpha\phi)=d\alpha\,\phi+(-1)^{\deg(\alpha)}\alpha\nabla\phi
$$
The {\it curvature\/} of $\nabla$ is then $R(\nabla)
=-\nabla^2:E\to\Omega^2\ot_AE$,
and obeys $-\nabla^2(f\,s)=f(-\nabla^2)s$ for any $f\in A$ and $s\in E$.

Suppose now that $\Omega^\cdot$ is involutive, i.e.\ there is
an antilinear antiautomorphism $\alpha\mapsto\alpha^*$ with
$\alpha^{**}=\alpha$, for all $\alpha\in \Omega^\cdot$.
Assume that $\deg(\alpha^*)=\deg(\alpha)$ and \hfill\break $(d\alpha)^*
=(-1)^{\deg(\alpha)+1}
d(\alpha^*)$
for homogeneous $\alpha$.
If $A$ is any involutive algebra, then the algebra $\Omega^\cdot(A)$
of universal differential forms is involutive, with the above
properties, if we set $(df)^*=-d(f^*)$ for $f\in A$.
In general, elements of $A$ of
the form $g=f^*f$, are called non-negative ($g\geq 0$).
The module $E$ is called {\it hermitian\/} if it has a
hermitian inner product
$(\, ,\,):E\times E\to A$, which is by definition
a sesquilinear form such that

\item{(i)}$ (f\,s,g\,t)=f\,(s,t)\,g^*,\qquad f,g\in A,\quad s,t\in E$

\item{(ii)} $(s,s)\geq 0$.
\item{(iii)} The map $s\mapsto (s,\cdot)$ from $E$ to the left
$A$ module
$E^*=\{l:E\to A,\quad l(f\,s+g\,t)=l(s)f^*+l(t)g^*\}$ is an isomorphism.

\noindent Any hermitian inner product on $E$ extends uniquely
to a sesquilinear map $\Omega^\cdot E\times \Omega^\cdot E
\to \Omega^\cdot$ such that $(\alpha\phi,\beta\psi)=
\alpha(\phi,\psi)\beta^*$ for all $\alpha$, $\beta\in\Omega^\cdot$,
$\phi$, $\psi\in \Omega^\cdot E$.
A connection $\nabla$ on a hermitian $A$-module $E$ is
{\it unitary\/} if, for all $s$, $t\in E$,
$ d(s,t)=(\nabla s,t)-(s, \nabla t)$ (the minus sign appears here
because we have set $(df)^*=-d\,f^*$). One has then for
homogeneous $\phi$, $\psi\in \Omega^\cdot E$
$$
d(\phi,\psi)=(\nabla\phi,\psi)-(-1)^{\deg(\phi)\deg(\psi)}(\phi,\nabla\psi)
$$

Let us now suppose that $\Omega^\cdot$ is an algebra over
$\R$, and take $E$ to be $\Omega^1$. This is the setting of
Riemannian geometry.

The {\it torsion\/} of a connection $\nabla:\Omega^1\to\Omega^1
\ot\Omega^1$ is
$$
T(\nabla)=d-m\circ\nabla
$$
It is an $A$ linear operator from $\Omega^1$ to $\Omega^2$.
The connections of interest in Riemannian geometry are those with vanishing
torsion. Among these connections we should like to find ones that can be
interpreted as natural generalizations of {\it Levi-Civita connections}. This
suggests to introduce the notion of a {\it metric} in non-commutative geometry.
One might think that a metric is specified by an inner product on $\Omega^1$.
However, in general, it does not appear to make sense to demand this inner
product to be hermitian. Actually, it looks more promising to first introduce
the notion of a {\it distance} on a non-commutative space, as proposed by
Connes and Lott [3].

Apparently, in non-commutative geometry the natural notion
of distance is provided by $K$-cycles. Recall that a
 $K$-cycle over an involutive algebra $A$
is a pair $(H,D)$, where $H=H_+\oplus H_-$ is
a $\Z_2$ graded Hilbert space with a $*$-action of $A$ by even
bounded operators, and $D$ is a possibly unbounded, odd
self-adjoint operator, called Dirac operator,
such that $[D,f]$ is bounded for
all $f\in A$ and $(D^2+1)^{-1}$ is compact.
Then $\pi(f_0df_1\cdots df_n)=f_0[D,f_1]\cdots[D,f_n]$ defines
an involutive (i.e.\ with $\pi(\alpha^*)=\pi(\alpha)^*$)
representation of the algebra $\Omega^\cdot(A)$ of
universal forms. One shows then that the graded subcomplex
$\Ker(\pi)+d\,\Ker(\pi)$ is a two-sided ideal of $\Omega^\cdot(A)$,
so that the quotient
$$
\Omega_D^\cdot(A)=\Omega^\cdot(A)/ \bigl( \Ker(\pi)+d\,\Ker(\pi) \bigr)
$$
is a graded differential algebra\footnote{${}^1$}{This algebra
was introduced in the Carg\`ese lecture
notes of Connes and Lott [3]. It replaces
the algebra of universal forms used in [1,2] and allows for
a more transparent treatment of ``auxiliary fields''.}.

A {\it Riemannian metric} is a hermitian inner product $-$ more generally, a
non-degenerate inner product $-$ on $\Omega_D^1 \, \equiv \, \Omega_D^1 (A)$
which (in the examples considered below) determines a notion of {\it distance}
coinciding with the one obtained from the Dirac operator, as in [3]. A
connection $\nabla$ on $\Omega^1$ is a {\it Levi-Civita}
connection if it has vanishing torsion and if $\pi (\nabla )$ is
{\it unitary} with respect to the metric on $\Omega_D^1$. In general it is not
true, as it is in the classical case, that for an arbitrary Riemannian metric
there is precisely one Levi-Civita connection.

It is straightforward to derive {\it Cartan structure equations} in this
context. Suppose that $\Omega_D^1$ is a trivial vector bundle, i.e. a free,
finitely generated $A$ module, with Riemannian metric. (The following analysis
could be generalized to situations where $\Omega_D^1$ is a non-trivial vector
bundle by introducing a suitable family of $\Omega_D^1 -$invariant subspaces
of $H$, with the property that the restriction of $\Omega_D^1$ to every
subspace in this family is trivial.) Let $E^A , A \, = \, 1, \dots , N$, be a
basis of sections of $\Omega_D^1$ which is orthonormal in the metric on
$\Omega_D^1$. We define $\Omega^{AB} \in \Omega_D^1$ by
$$
\pi (\nabla ) E^A \enskip = \enskip - \, \sum_B \, \Omega^{AB} \,
\otimes \, E^B \enskip .
$$
The components of torsion and curvature are defined by
$$
\eqalign{
\pi \biggl( T (\nabla ) \biggr) \, E^A \enskip &= \enskip T^A \cr
\pi \biggl( R (\nabla ) \biggr) \, E^A \enskip &= \enskip \sum_B \,
R^{AB} \, \ot \, E^B \enskip . \cr }
$$
The Cartan structure equations follow by inserting the definitions of $T
(\bigtriangledown )$ and $R (\bigtriangledown)$ :
$$
\eqalign{
T^A \enskip &= \enskip \pi (d \tilde{E}^A ) \enskip + \enskip \sum_B \,
\Omega^{AB} E^B \enskip , \cr
R^{AB} \enskip &= \enskip \pi (d \tilde{\Omega}^{AB}) \enskip + \enskip \sum_C
\, \Omega^{AC} \Omega^{CB}, \cr }
$$
where $\tilde{E}^A , \tilde{\Omega}^{AB}$ are representatives of $E^A ,
\Omega^{AB}$, respectively, in $\Omega^1$.

\def\Cliff{{\rm Cliff}}
We now introduce a class of algebras and of $K$-cycles for
which the Riemannian geometry concepts introduced above can
be defined.

Let $X$ be a compact even dimensional
$C^\infty$ spin manifold, with a reference
Riemannian metric $g_0$ and fixed spin structure,
$A$ the algebra of smooth
real functions on $X$ and $\Cliff(T^*X)$ the Clifford bundle over $X$,
whose fiber at $x$ is the (real) Clifford algebra of the cotangent
space $\Cliff(T^*_xX)$ associated to $g_0(x)$. Let $S$ be the spinor bundle.
Thus $S$ is a $\Z_2$ graded complex vector
bundle over $X$, with a representation of
the Clifford algebra of the cotangent space on each fiber $S_x$,
such that ${\rm End}_{\C}(S_x)\simeq \Cliff(T_x^*X)\otimes\C$.
A section of $\End(S)\simeq \Cliff(T^*X)\otimes\C$ is called
real if it takes values in the real Clifford algebra.
We consider $K$-cycles $(H,D)$ where:

\item{(i)} $D$ is an odd first order elliptic differential operator
on the space $C^\infty(S)$ of smooth sections of $S$.
\item{(ii)} For each $f\in A$, $[D,f]$ is a real section
of ${\rm End}(S)$.
\item{(iii)}
$H=L^2(S,\rho d^4 y)$ is the space of square integrable
sections of $S$, where $\rho(y)$ is a density for
which $D$ is self-adjoint.

We will also need the following variant with group action:
Let $X$, $A$, $\Cliff(T^*X)$ and $S$ be as above, and suppose
that $X$ is a finite smooth covering of a manifold $Y$. That
is, $p:X\to Y$ is a principal $G$ bundle with base space $Y=X/G$, and
$G$ is a finite group. The reference metric will be chosen
to be preserved by the group action, and we assume that
the group action lifts to $S$. Denote by $p_*S$ the
vector bundle over $Y$ whose fiber over $y$ is the
direct sum $\oplus_{p(x)=y}S_x$. Both $A$ and the
group $G$ act on the sections of $p_*S$. A linear
operator on the space of smooth section of $p_*S$ is
called equivariant if it commutes with the action of $G$.
The vector space ${\rm End_{\C}}(p_*S_y)$ is the space
of matrices indexed by $p^{-1}(y)$ with entries in
$\Cliff(T_y^*Y)\otimes\C$. A vector in ${\rm End_{\C}}(p_*S_y)$
is called real if its matrix entries belong to the
real Clifford algebra, and a section of ${\rm End}(p_*S)$ is
called real if it takes real values.

In this setting, we consider $K$-cycles $(H,D)$ where:

\item{(i)}  $D$ is an odd equivariant
 first order elliptic differential operator
on the space $C^\infty(p_*S)$ of smooth sections of $p_*S$.
\item{(ii)} For each $f\in A$, $[D,f]$ is multiplication
by a real section of ${\rm End}(p_*S)$
\item{(iii)}
$H=L^2(p_*S,\rho d^4y)$ is the space of square integrable
sections of $p_*S$, where $\rho(y)$ is a density for
which $D$ is self-adjoint.

These data define a Riemannian geometry on the
graded differential algebra $\Omega^*_D(A)$. The
Riemannian metric is defined to be
$$
G(\alpha,\beta)=\tr(\pi(\alpha^*)\pi(\beta)),
\qquad \alpha,\beta\in\Omega^1_D(A).
$$
This is independent of the choice of representatives $\alpha$,
$\beta$ since $d\Ker(\pi)\cap\Omega^1(A)=0$ and therefore
$\Omega^1_D(A)$ is isomorphic to $\pi(\Omega^1(A))$.
The trace over the Clifford algebra
is defined fiberwise. We normalize it in
such a way that the trace of the identity is one.

\vfill\eject


{\bf\noindent 3. A gravity action}

Let us apply the formalism introduced in the previous section
to an example. As in [1--3], we take $X$ to be two copies
of a compact, say four-dimensional, spin manifold $Y$:
$$X=Y\times \Z_2,$$
and we have the trivial $\Z_2$ bundle $p:X\to Y$.
The algebra $A$ is then $C_{\R}^\infty(Y)\oplus C_{\R}^\infty(Y)$.
It is convenient to think of $A$ as a subalgebra of diagonal
matrices in the algebra $M_2(\C)\otimes C^\infty(\Cliff(T^*Y))$ of two by two
matrices whose entries are smooth sections of the Clifford bundle. The
chirality operator $\gamma^5$  belongs to the real Clifford algebra and
defines a $\Z_2$ grading of the spinor bundle $S$. The operator
$$
\Gamma=\left(\matrix{\gamma^5 & 0\cr 0 & -\gamma^5}\right)
$$
defines a $\Z_2$ grading of $C^\infty(p_*S)=C^\infty(S)\oplus
C^\infty(S)$ (the minus sign is a matter of convention).

We work in local coordinates. Let us introduce gamma matrices
$\gamma^a$ with $(\gamma^a)^*=-\gamma^a$, $a=1,\dots,4$,
obeying the relations $\gamma^a\gamma^b+\gamma^b\gamma^a=
-2\delta^{ab}$. Then $\gamma^5=\gamma^1\gamma^2\gamma^3\gamma^4$
is self-adjoint and has square one. We set
$\gamma^{ab}={1\over2}(\gamma^a\gamma^b-\gamma^b\gamma^a)=
-(\gamma^{ba})^*$.

The Dirac operator can then be represented as a two by two
matrix $(D_{ij})$, $i,j\in\{+,-\}$, whose entries are
first order differential operators acting on spinors of $Y$.
What are the restrictions on these entries imposed
by (i)--(iii)? First of all, $\Z_2$ equivariance implies
that $D_{+-}=D_{-+}$ and $D_{++}=D_{--}$, and the fact that
$[D,f]$ is a multiplication operator implies that
$D_{+-}$ should be a multiplication operator.
The most general form of $D$, compatible with self-adjointness,
reality and oddness is then
$$
D=\left(\matrix
{\gamma^a\epsilon_a^\mu\partial_\mu+\cdots &
\psi+\gamma^5\phi \cr
\psi+\gamma^5\phi &
\gamma^a\epsilon_a^\mu\partial_\mu+\cdots}
\right)
$$
where $\epsilon_a^\mu$, $\psi$ and $\phi$ are real functions.
Since $D$ is elliptic, $\epsilon_a^\mu\partial_\mu$
is a basis of the tangent space, and we can define
a Riemannian metric $g$ on $Y$ by $g(\epsilon_a,\epsilon_b)
=\delta_{ab}$. The dots in the definition of $D$
indicate zero order contributions which do not contribute
to $\pi$.

\def\tr{{\rm tr}}
\def\Tr{{\rm Tr}}
The representation $\pi$ on one-forms can now be computed.
 Let $\alpha=\Sigma_ia_idb_i\in\Omega^1(A)$
be a representative of  a one-form in $\Omega^1_D(A)$.
Then $\pi(\alpha)$ is parametrized by two classical
one-forms $\alpha_{1\mu}$, $\alpha_{2\mu}$, and two
functions $\alpha_5$,
$\tilde\alpha_5$, on $Y$:
$$
\pi(\alpha)=
\left(
\matrix{
\gamma^\mu\alpha_{1\mu} & \bar{\gamma}\alpha_5\cr
-\bar{\gamma}\tilde\alpha_5 &\gamma^\mu\alpha_{2\mu}}  \right).
$$
We use the notation $\gamma^\mu=\gamma^a \varepsilon_a^\mu$,
$\bar{\gamma}=\psi+\gamma^5\phi$. In terms of the variables
$a_i=a_{i1}\oplus a_{i2}$ and
$b_i=b_{i1}\oplus b_{i2}$, we have
$$
\eqalign
{
\alpha_{1\mu}&=\sum_ia_{i1}\partial_\mu b_{i1},\cr
\alpha_{2\mu}&=\sum_ia_{i2}\partial_\mu b_{i2},\cr
\alpha_5&=\sum_ia_{i1}(b_{i2}-b_{i1}),\cr
\tilde\alpha_5&=\sum_ia_{i2}(b_{i2}-b_{i1}),}
$$
The Riemannian metric $G:\Omega_D^1(A)\otimes
\Omega_D^1(A)\to A$ can be expressed, using the isomorphism
$\Omega_D^1(A)=\pi(\Omega^1(A))$,
in terms of components:
$$
G(\alpha,\beta)
=(g^{\mu\nu}\alpha_{1\mu}\beta_{1\nu}+g^{55}\tilde\alpha_5\tilde\beta_5)
\oplus
(g^{\mu\nu}\alpha_{2\mu}\beta_{2\nu}+g^{55}\alpha_5\beta_5),
$$
where $g^{\mu\nu}=-\tr(\gamma^\mu\gamma^\nu)
=\epsilon^\mu_a\epsilon^\nu_a$ and
$g^{55}=\tr\bar{\gamma}^2=\psi^2+\phi^2$.

To compute torsion and curvature, we must understand two-forms,
$\Omega_D^2(A)$. This space is isomorphic to the quotient
of $\pi(\Omega^2(A))$ by the space of
``auxiliary fields'' \hfill\break
$\pi(d\,\Ker(\pi|_{\Omega^1(A)}))$.
We proceed to compute the general form of auxiliary
fields. If $\alpha=\Sigma_ia_idb_i\in\Ker(\pi)$,
we obtain for $\pi(d\alpha)=\Sigma_i[D,a_i][D,b_i]$,
$$
\pi(d\alpha)=\left(\matrix{
-g^{\mu\nu}\partial_\mu a_{i1}\partial_\nu b_{i1} &
-2\psi\gamma^\mu a_{i1}\partial_\mu b_{i2} \cr
-2\psi\gamma^\mu a_{i2}\partial_\mu b_{i1} &
-g^{\mu\nu}\partial_\mu a_{i2}\partial_\nu b_{i2}
}
\right),
$$
and it is not difficult to see that, for a suitable
choice of $a_i$, $b_i$ subject to the constraint
$\pi(\alpha)=0$, any expression of the form
$$
\left(
\matrix{X_1 & \psi\gamma^\mu Y_\mu\cr
\psi\gamma^\mu\tilde Y_\mu & X_2}
\right)
$$
can be obtained.

Next, we express $\pi(d\alpha)$ modulo auxiliary fields for
any one-form $\alpha$ in terms of its components:
$$
\pi(d\alpha)=
\left(
\matrix
{\gamma^{\mu\nu}\partial_\mu\alpha_{1\nu}
+2\phi\psi\gamma^5(\alpha_5-\tilde\alpha_5)
&
\phi\gamma^\mu\gamma^5(\partial_\mu\alpha_5+\alpha_{1\mu}-\alpha_{2\mu})
\cr
-\phi\gamma^\mu\gamma^5(\partial_\mu\tilde\alpha_5
+\alpha_{1\mu}-\alpha_{2\mu})
&
\gamma^{\mu\nu}\partial_\mu\alpha_{2\nu}
+2\phi\psi\gamma^5(\alpha_5-\tilde\alpha_5)
}\right).
$$
This choice of representative in the class
of $\pi(d\alpha)$ in
$\pi(\Omega^2(A))/\pi(d\Ker(\pi|_{\Omega^1(A)}))$
is uniquely determined by the property to be
{\it orthogonal\/}
to all auxiliary fields, with respect to the inner product
on $\Omega^2(A)$ defined by the Dixmier trace:
$$
(\alpha,\beta)=\Tr_\omega(\pi(\alpha)^*\pi(\beta)
|D|^{-4}).
$$

For explicit calculations it is convenient to introduce
local orthonormal bases $\{E^A\}$ of $\Omega_D^1(A)$.
We use the following convention for indices: capital
letters $A$, $B$, \dots denote indices taking the values
1 to 5, and lower case letters $a$, $b$, \dots take values
from 1 to 4. Introduce a local  orthonormal frame of one-forms
$e^a_\mu dx^\mu$ on $Y$.
The basis is
$$
\eqalign{
E^a&=\left(\matrix{
\gamma^a & 0\cr
0 & \gamma^a}
\right)=\left(\matrix{
\gamma^\mu e^a_\mu & 0\cr
0 & \gamma^\mu e^a_\mu}
\right),\cr
E^5&=\left(\matrix{
0 & \bar{\gamma}\lambda\cr
-\bar{\gamma}\lambda & 0}
\right),\qquad \lambda=(\phi^2+\psi^2)^{-{1\over2}}.
}
$$

Suppose now that the connection $\nabla$ is unitary with respect to the
given $K$-cycle. The components of the one-form corresponding to $\pi (\nabla
)$ are denoted by
$$
\Omega^{AB}=
\left(\matrix{
 \gamma^\mu\omega_{1\mu}^{AB} & \bar{\gamma} \ell^{AB}
 \cr
 -\bar{\gamma}\tilde\ell^{AB} & \gamma^\mu\omega_{2\mu}^{AB}
 }
\right).
$$
The unitarity condition
$(\Omega^{AB})^*=\Omega^{BA}$
implies the component relations
$$\eqalign{
\omega_{1\mu}^{AB}&=-\omega_{1\mu}^{BA}\cr
\omega_{2\mu}^{AB}&=-\omega_{2\mu}^{BA}\cr
\tilde\ell^{AB}&=-\ell^{BA}}
$$
The components of torsion and curvature
are readily computed. As above, we give the representative
in $\Omega^2_D(A)$ orthogonal to auxiliary fields.
For the torsion we find
$$
\eqalign{
T^a&=
\left(\matrix{
\gamma^{\mu\nu}
(\partial_\mu e^a_\nu+\omega_{1\mu}^{ab}e^b_\nu)
-2\phi\psi\lambda\gamma^5\ell^{a5}
&
-\phi\gamma^\mu\gamma^5
(\ell^{ab}e^b_\mu-\lambda\omega_{1\mu}^{a5})
 \cr
 \phi\gamma^\mu\gamma^5
(\tilde\ell^{ab}e^b_\mu-\lambda\omega_{2\mu}^{a5})
     &
\gamma^{\mu\nu}
(\partial_\mu e^a_\nu+\omega_{2\mu}^{ab}e^b_\nu)
-2\phi\psi\lambda\gamma^5\tilde\ell^{a5}
}
\right),
\cr
\
\cr
T^5&=
\left(\matrix{
\gamma^{\mu\nu}
\omega_{1\mu}^{5b} e^b_\nu
-2\phi\psi\lambda\gamma^5\ell^{55}
&
\phi\gamma^\mu\gamma^5
(\partial_\mu\lambda-\ell^{5b}e^b_\mu) \cr
-\phi\gamma^\mu\gamma^5
(\partial_\mu\lambda-\tilde\ell^{5b}e^b_\mu)      &
\gamma^{\mu\nu}
\omega_{2\mu}^{5b} e^b_\nu
+2\phi\psi\lambda\gamma^5\ell^{55}
}
\right).
}
$$
The expression for the curvature is
$$
R^{AB}=
\left(\matrix{
\gamma^{\mu\nu}
R_{1\mu\nu}^{AB}
+2\phi\psi\gamma^5 P_1^{AB}
&
\phi\gamma^\mu\gamma^5
Q_\mu^{AB}\cr
-\phi\gamma^\mu\gamma^5
\tilde Q_\mu^{AB}
&
\gamma^{\mu\nu}
R_{2\mu\nu}^{AB}
+2\phi\psi\gamma^5 P_2^{AB}
}
\right),
$$
where
$$
\eqalign
{
R_{i\mu\nu}&=
\partial_\mu\omega_{i\nu}^{AB}
-\partial_\nu\omega_{i\mu}^{AB}
+
\omega_{i\mu}^{AC}\omega_{i\nu}^{CB}
-
\omega_{i\nu}^{AC}\omega_{i\mu}^{CB},\qquad i=1,2,\cr
Q_\mu^{AB}&=
\partial_\mu\ell^{AB}+
\omega_{1\mu}^{AB}-\omega_{2\mu}^{AB}
+\omega_{1\mu}^{AC}\ell^{CB}-\omega_{2\mu}^{CB}\ell^{AC},\cr
\tilde Q_\mu^{AB}&=
-\partial_\mu\ell^{BA}+
\omega_{1\mu}^{AB}-\omega_{2\mu}^{AB}
+\omega_{1\mu}^{CB}\ell^{CA}-\omega_{2\mu}^{AC}\ell^{BC},\cr
P_1^{AB}&=\ell^{AB}+\ell^{BA}+\ell^{AC}\ell^{BC},\cr
P_2^{AB}&=\ell^{AB}+\ell^{BA}+\ell^{CA}\ell^{CB},}
$$
As a gravity action we propose the following generalized Einstein-Hilbert
action, given in terms of the inner product
$(\alpha,\beta)=\Tr_\omega(\pi(\alpha)^*\pi(\beta)|D|^{-4})$ on
two-forms defined through the identification
of $\Omega^2_D(A)$ with  $\pi(\Omega^2(A))
\cap\pi(d\Ker(\pi|_{\Omega^1(A)})^\perp$):
$$
I=(E^AE^B,R^{AB})
$$
This action reduces to (and could be alternatively defined as)
the integral over $Y$,
$$
I=\int_Y\tr\bigl( (E^A E^B)^* R^{AB}\bigr)\sqrt g \, d^4 y.
$$
(Here the trace is over $\End(p_*S_y)$).
Inserting the above expressions for $E^A$ and
$R^{AB}$ yields the action as a function of the component fields.
Set $U^a_\mu=
Q^{a5}_\mu+\tilde Q^{a5}_\mu-
Q^{5a}_\mu-\tilde Q^{5a}_\mu$. The result is
$$
I=\int_Y [
\epsilon_a^\mu\epsilon_b^\nu
(R_{1\mu\nu}^{ab}+R_{2\mu\nu}^{ab})
+
\lambda\phi^2
\epsilon_a^\mu U_\mu^a
-4\phi^2\psi^2\lambda^2(P_1^{55}+P_2^{55})]\sqrt g \, d^4 y.
$$
At this point two possibilities are open. One can either
take the action $I$ as a starting point, with all fields
independent, and eliminate non-dynamical fields by
their equations of motion. Or one can impose the
torsion constraint, and derive an action for
Levi-Civita connections. We will follow the second
approach.

It turns out that, in general, one gets an
uninteresting model, describing just two decoupled
universes. A more interesting example is obtained by
imposing the additional condition $\psi=0$.
In other words, we consider on Dirac operators
of the form
$$
D=
\left(\matrix{
\gamma^a e^\mu_a\partial_\mu+\cdots & \gamma^5\phi(x) \cr
\gamma^5\phi(x) & \gamma^a e^\mu_a\partial_\mu+\cdots
}
\right),
$$
which is in fact closer to the form of Dirac operators
used in particle models [1--6].

The zero torsion condition has the following consequences
for the components:

\noindent 1. $\omega^{ab}_\mu\equiv\omega^{ab}_{1\mu}=\omega^{ab}_{2\mu}$ is
the one form corresponding to the classical Levi-Civita connection
of the metric $g_{\mu\nu}=e^a_\mu e^a_\nu$. It
is the unique solution of
$
de^a+\omega^{ab}\wedge e^b=0$,
$\omega^{ab}=-\omega^{ba}$

\noindent 2. $\ell^{ab}=\ell^{ba}$, $\ell^{5a}=-\ell^{a5}$.

\noindent 3. $\omega_{1\mu}^{a5}=-\omega_{2\mu}^{a5}=\lambda^{-1}
\ell^{ab}e^b_\mu$

\noindent 4. $\partial_\mu\lambda=e^a_\mu\ell^{5a}$

It is interesting to notice that the zero torsion constraint
selects $\Z_2$-equivariant connections. In other words, let
$\theta:A\to A$ be the involution $\theta(a_1\oplus a_2)=
a_2\oplus a_1$. Extend it, using the equivariance of $D$,
 to the unique involutive
automorphism of $\Omega^\cdot_D(A)$ such that $d\theta=\theta d$.
Then Levi-Civita connections have the property $\theta\otimes\theta
\nabla=\nabla\theta$.

The resulting gravity action is then
$$
I=\int_Y[2R-\lambda^{-1}4\nabla_\mu\partial^\mu\lambda
+4\lambda^{-2}\ell^{aa}\ell^{55}
+\lambda^{-2}(\ell^{aa}\ell^{bb}-\ell^{ab}\ell^{ab})]
\sqrt gd^4y
$$
The fields $\ell^{ab}$, $\ell^{55}$ decouple, and with
the substitution $\lambda={\rm exp}(\sigma)$,
we finally obtain the action of a massless
scalar coupled to the gravitational field:
$$
I=2\int_Y[R-2\partial_\mu\sigma\partial^\mu\sigma]\sqrt gd^4x.
$$

To understand the role of the field $\sigma$ we can study the coupling of
gravity to the Yang-Mills sector. In particular, in the example of the
standard model in [3] we see that $g^{\mu \nu }$ is the metric of the
Riemannian manifold while $\phi \, = \, e^{-\sigma}$ replaces the electroweak
scale $\mu$. In other words, the vacuum expectation value of the field
$\phi$ determines the electroweak scale, thus forming a connection between
gravity and the standard model. From the form of the gravity action, it is
clear that the field $\sigma$ has no potential. The only other term we could
have added is a cosmological constant
$$
(E^A E^B , \enskip E^A E^B )
$$
and this is $\sigma -$independent. This implies that at the classical level
the vacuum expectation value of $\phi$ is undetermined. It is conceivable that
the gravity action acquires a Coleman-Weinberg potential through quantum
effects. However, at present this is beyond our capabilities, since the problem
of quantization in non-commutative geometry has not as yet been dealt with.

\vskip.2truecm

\noindent {\bf Acknowledgements}

\vskip.2truecm

We thank {\sl A. Connes} and {\sl J. Lott} for very helpful discussions and
for providing us with an advance copy of [3].

\vfill
\eject

{\bf \noindent References}
\vskip.2truecm \frenchspacing

\item{[1]} A. Connes, in {\sl The interface of mathematics
and particle physics }, Clarendon press, Oxford 1990, Eds
D. Quillen, G. Segal and  S. Tsou

\item{[2]} A. Connes and J. Lott, {\sl Nucl. Phys. B Proc. Supp.}
{\bf 18B} 29 (1990), North-Holland, Amsterdam.

\item{[3]} A. Connes and J. Lott, The metric aspect of
noncommutative geometry, {\sl to appear in Proceedings of the
1991 Carg\`ese summer school}.

\item{[4]} D. Kastler , Marseille preprints

\item{[5]} R. Coquereaux, G. Esposito-Far\'ese, G. Vaillant,
{\sl Nucl. Phys.}{\bf B353} 689 (1991);\br
M. Dubois-Violette, R. Kerner, J. Madore, {\sl J. Math.
Phys.}{\bf 31} (1990) 316;\br
B. Balakrishna, F. G\"ursey and K. C. Wali, {\sl Phys. Lett.}
{\bf 254B} (1991) 430.

\item{[6]} A. H. Chamseddine, G. Felder and J. Fr\"ohlich,
Grand unification in non-commutative geometry, Z\"urich, preprint 1992.
\end